# A relativistic quantum theory of dyons wave propagation

B.C. Chanyal

**Abstract:** Beginning with the quaternionic generalization of the quantum wave equation, we construct a simple model of relativistic quantum electrodynamics for massive dyons. A new quaternionic form of unified relativistic wave equation consisting of vector and scalar functions is obtained, and also satisfy the quaternionic momentum eigenvalue equation. Keeping in mind the importance of quantum field theory, we investigate the relativistic quantum structure of electromagnetic wave propagation of dyons. The present quantum theory of electromagnetism leads to generalized Lorentz gauge conditions for the electric and magnetic charge of dyons. We also demonstrate the universal quantum wave equations for two four-potentials as well as two four-currents of dyons. The generalized continuity equations for massive dyons in case of quantum fields are expressed. Furthermore, we concluded that the quantum generalization of electromagnetic field equations of dyons can be related to analogous London field equations (i.e., current to electromagnetic fields in and around a superconductor).

*Key words:* quaternion, dyons, Lorentz gauge, relativistic quantum equations, wave propagation.

**Résumé :** Débutant avec la généralisation en quaternions de l'équation d'onde quantique, nous construisons un modèle simple de l'électrodynamique quantique pour des dyons massifs. Nous obtenons une nouvelle forme en quaternions de l'équation d'onde quantique relativiste, consistant de fonctions scalaires et vectorielles, qui satisfait aussi l'équation de valeurs propres d'impulsion relativiste des quaternions. Gardant en tête l'importance de la théorie quantique des champs, nous étudions la structure relativiste quantique de la propagation d'ondes électromagnétiques des dyons. L'actuelle théorie de l'électromagnétisme quantique mène à une condition de jauge de Lorentz généralisée pour les charges électrique et magnétique du dyon. Nous démontrons aussi l'équation d'onde quantique universelle pour deux 4-potentiels et deux 4-courants de dyons. Nous présentons l'équation de continuité généralisée pour des dyons massifs dans le cadre de champs quantiques. De plus, nous concluons que la généralisation quantique des équations de champ électromagnétiques peut être reliée à un analogue des équations de champ de London i.e. le courant d'un champ électromagnétique dans et autour d'un supraconducteur. [Traduit par la Rédaction]

*Mots-clés :* quaternion, dyon, jauge de Lorentz, équation quantique relativiste, propagation d'onde.

## 1. Introduction

Relativistic quantum theory is applicable for massive particles propagating at all velocities up to those comparable to the speed of light, and can accommodate massless particles. The theory has application in high-energy physics [1], particle physics and accelerator physics [2], atomic physics [3], and condensed matter physics. The relativistic formulation is more successful than the original quantum mechanics in some contexts, in particular: the prediction of antimatter, electron spin, spin magnetic moments of elementary spin-1/2 fermions, fine structure, and quantum dynamics of charged particles in electromagnetic fields [3]. For massive particles the classical description is given by Hamilton's variational principle and the nonrelativistic quantum mechanical description, by the Schrödinger equation. This equation represents the first quantization procedure wherein the particle shows the wave nature and the physical quantities like position, momentum, etc., are elevated to the status of operators. For electromagnetic fields, Maxwell equations are considered to be the classical description and the quantization of these fields gives the relativistic quantum description. Bialynicki-Birula [4, 5] and Sipe [6] explained that Maxwell equations are the quantum description of electromagnetic fields at the first quantum level and the Fermat principle for light forms the classical description of electromagnetic fields similar to Hamilton's variational principle for particles with mass.

However, Dirac [7] demonstrated quantum mechanics apparently predicts that a magnetic charge, if it is ever found in nature, must be quantized in units of $\hbar c/2e$ where $e$ is the absolute electron charge value. This result has been considered one of the most remarkable predictions of quantum mechanics, which has yet to be verified experimentally [8]. Dirac [9] also developed a general dynamical theory of the magnetic monopole reconfirming and extending his original results. Keeping in mind 't Hooft's solutions [10] and the fact that despite the potential importance of monopoles, the formalism necessary to describe them has been clumsy and not manifestly covariant, Rajput et al. [11] developed the self-consistent quantum field theory of generalized electromagnetic fields associated with dyons (particles carrying electric and magnetic charges).

In general, there are a number of requirements that need to be imposed on a formalism for a quantum mechanical description of electromagnetic fields of dyons. For example, the time-dependent solutions of the Maxwell equations for dyons provide the basis for both classical electromagnetic theory and quantum electrodynamics. Another way is that the wave functions obey the quantum mechanical principle of linear superposition as well as the formalism being Lorentz invariant to properly describe the space–time behavior of the electromagnetic radiation in presence of dyons. Furthermore, in quantum formulation we also can take an approximation scheme like the reduction of the Dirac equation to obtain the Schrödinger equation for dyons. Therefore, we can say that it is necessary for the formalism to provide the tools for methods associated with quantum mechanics. In this paper, we present a quantum analogue of classical field equations of dyons in terms of four-dimensional quaternion formalism. Starting with the quaternionic generalization of the quantum wave equation, we construct a theory of relativistic quantum electrodynamics for massive dyons having both electric and magnetic charges. The present quantum theory leads to generalized Lorentz gauge conditions, respectively, for the electric charge and magnetic monopoles of dyons. As such we expressed the generalized potential wave equations, current wave equation, and continuity equations for massive dyons in the case of quantum fields. Accordingly, we also describe the quantum wave propagation in an external electromagnetic field of dyons. Furthermore, the generalized quantum equations of dyons can be used to show an analogue of the London field equations in the case of a superconductor.

There has been a revival in the formulation of natural laws within the framework of general quaternion algebra and basic physical equations. Quaternions [12] were the very first example of hyper-complex numbers having significant impacts on mathematics and physics. The quaternion algebra $\mathbb{Q}$ is a four-dimensional algebra [12, 13] over the field of real numbers, $\mathbb{R}$. The quaternion is very suitable to express the typical four-qualities in physics, such as four-position, four-momentum, four-force, four-potential, and four-current. Similar to a complex number $\mathbb{C} = q_0 + iq_1$ where $i^2 = -1$ and $(q_0, q_1) \in \mathbb{R}$, a real quaternion $\mathbb{Q}$ and its conjugate $\overline{\mathbb{Q}}$ may be written in terms of scalar $q_0$ and vector $\boldsymbol{q}$, that is,

$$\mathbb{Q} = (q_0, \boldsymbol{q}) \qquad \overline{\mathbb{Q}} = (q_0, -\boldsymbol{q}) \tag{1}$$

The real part of the quaternion $q_0$ is defined by $\mathfrak{Re}\mathbb{Q} = (\overline{\mathbb{Q}} + \mathbb{Q})/2 \to q_0$, and the imaginary part is called pure quaternion written by $\mathfrak{Im}\mathbb{Q} = (\mathbb{Q} - \overline{\mathbb{Q}})/2 \to \{\boldsymbol{q} = q_j e_j, (j = 1, 2, 3)\}$ where $e_j$ are quaternion units. The quaternion units $(e_0, e_1, e_2, e_3)$ are known as the quaternion basis and satisfy the following conditions:

$$\begin{aligned} e_0^2 = e_0 = 1 \quad e_j^2 = -e_0 \quad e_0 e_j = e_j e_0 = e_j \quad (j = 1, 2, 3) \\ e_i e_j = -\delta_{ij} + \varepsilon_{ijk} e_k \quad (\forall\, i, j, k = 1, 2, 3) \end{aligned} \tag{2}$$

Here $\delta_{ij}$ is the delta symbol and $\varepsilon_{ijk}$ is the Levi–Civita three index symbol having value ($\varepsilon_{ijk} = +1$) for cyclic permutation, ($\varepsilon_{ijk} = -1$) for anti-cyclic permutation and ($\varepsilon_{ijk} = 0$) for any two repeated indices. Addition and multiplication are defined by the usual distribution law $(e_j e_k)e_l = e_j(e_k e_l)$ along with the multiplication rules given by (2). We know that $\mathbb{Q}$ is an associative but non-commutative algebra. If $q_0, q_1, q_2, q_3$ are taken as complex quantities, the quaternion is said to be a bi-quaternion or complex quaternion. In practice $\mathbb{Q}$ is often represented as a 2 × 2 matrix so that $\mathbb{Q} = q_0 - i(\boldsymbol{\sigma} \cdot \boldsymbol{q})$ where $e_0 = I_{2\times 2}$, $e_j = -i\sigma_j$ ($j = 1, 2, 3$) and $\sigma_j$ are the usual Pauli-spin matrices. The internal and external vector products and the vector itself can be used in quaternion equations.

## 2. Electromagnetic field equations of dyons

In special relativity, a four-vector has four components, which transform in a specific way under Lorentz transformations. Specifically, a four-vector is an element of a four-dimensional vector space considered as a representation space of the standard representation of the Lorentz group. The description of phenomena at high energies requires the investigation of relativistic wave equations. To write the relativistic phenomena of dyons, let us start with the four-vector representation along with an imaginary fourth component of Euclidean structure (+, +, +, −) as

- world vector

$$\mathbb{X} = \{x, y, z, -ict\} \tag{3}$$

- four-gradiant

$$\Box = \left\{\frac{\partial}{\partial x}, \frac{\partial}{\partial y}, \frac{\partial}{\partial z}, -\frac{\partial}{i\partial(ct)}\right\} \tag{4}$$

- four-momentum

$$\mathbb{P} = \left\{p_x, p_y, p_z, -\frac{i}{c}E\right\} \tag{5}$$

- electric four-potential

$$\mathbb{A} = \left\{A_x, A_y, A_z, -\frac{i}{c}\phi_e\right\} \tag{6}$$

- magnetic four-potential

$$\mathbb{B} = \left\{B_x, B_y, B_z, -\frac{i}{c}\phi_m\right\} \tag{7}$$

- electric four-current

$$\mathbb{J} = \{J_x, J_y, J_z, -ic\rho_e\} \tag{8}$$

- magnetic four-current

$$\mathbb{K} = \{K_x, K_y, K_z, -ic\rho_m\} \tag{9}$$

where the letters $\mathbb{X}, \Box, \mathbb{P}, \mathbb{A}, \mathbb{B}, \mathbb{J}, \mathbb{K}$ abbreviate the full four-vector. As such, we can write the quaternionic four-vector in terms of covariant form

$$\mathbb{X}^\nu = \{x^0, x^1, x^2, x^3\} = (x^0 e_0 + x^1 e_1 + x^2 e_2 + x^3 e_3) \qquad (\mathbb{X}^\nu \in \mathbb{Q}) \tag{10}$$

and

$$\sum_{\nu=0}^{3} \mathbb{X}^\nu \overline{\mathbb{X}}_\nu \equiv (x^0 x_0 + x^1 x_1 + x^2 x_2 + x^3 x_3) = -c^2 t^2 + x^2 + y^2 + z^2 \tag{11}$$

where $\overline{\mathbb{X}}_\nu$ is the quaternion conjugate of $\mathbb{X}_\nu$. The d'Alembert operator, defined by $\Box$, can be expressed as

$$\Box = \Box \cdot \overline{\Box} = \frac{\partial^2}{\partial x^2} + \frac{\partial^2}{\partial y^2} + \frac{\partial^2}{\partial z^2} - \frac{1}{c^2}\frac{\partial}{\partial t^2} = \left(\nabla^2 - \frac{1}{c^2}\frac{\partial}{\partial t^2}\right) \tag{12}$$

Thus in this four-dimensional theory, a dyon is defined as a particle that simultaneously carries electric and magnetic charge. Many grand unified theories predict the existence of both magnetic monopoles and dyons. Starting with the idea of quantum electrodynamics given by Cabibbo and Ferrari [14], we may write the following complex quantity with their real and imaginary parts as electric and magnetic constituents of dyons, that is,

- generalized charge

$$Q\{e, g\} = e + ig \tag{13}$$

- generalized potential

$$\mathbb{V}^\nu\{A^\nu, B^\nu\} = A^\nu + iB^\nu \quad (14)$$

- generalized current

$$\mathbb{J}^\nu\{J^\nu, K^\nu\} = J^\nu + iK^\nu \quad (15)$$

where $i = \sqrt{-1}$ and $e, g, A^\nu, B^\nu, J^\nu$, and $K^\nu$ define the electric charge, magnetic charge, electric four-potential, magnetic four-potential, electric four-current, and magnetic four-current, respectively. We may write the symmetric electromagnetic field equations called generalized Dirac Maxwell's (GDM) equations for dyons as

$$\begin{cases} \nabla \cdot \mathcal{E} = \dfrac{\rho_e}{\varepsilon_0} \\ \nabla \cdot \mathcal{H} = \mu_0 \rho_m \\ \nabla \times \mathcal{E} = -\dfrac{\partial \mathcal{H}}{\partial t} - \mu_0 \mathbf{K} \\ \nabla \times \mathcal{H} = \dfrac{1}{c^2}\dfrac{\partial \mathcal{E}}{\partial t} + \mu_0 \mathbf{J} \end{cases} \quad (16)$$

Here $\mathcal{E}$ is the electric field, $\mathcal{H}$ is the magnetic field, $\rho_e$ is the charge source density due to electric charge, $\rho_m$ is the charge source density due to magnetic charge (monopole), $\mathbf{J}$ is the current source density due to electric charge, and $\mathbf{K}$ is the current source density due to magnetic charge. As such the electric and magnetic fields of dyons satisfying GDM equations (16) are now expressed in terms of the components of two four-potentials in a symmetrical manner, that is,

$$\mathcal{E} = -\nabla \phi_e - \frac{\partial \mathbf{A}}{\partial t} - \nabla \times \mathbf{B} \quad (17)$$

$$\mathcal{H} = -\nabla \phi_m - \frac{\partial \mathbf{B}}{\partial t} + \nabla \times \mathbf{A} \quad (18)$$

The vector field $\mathbf{\Omega}$ associated with generalized electromagnetic fields of dyons is defined as

$$\mathbf{\Omega} = \mathcal{E} + i\mathcal{H} = \left(-\nabla\phi_e - \frac{\partial \mathbf{A}}{\partial t} - \nabla \times \mathbf{B}\right) + i\left(-\nabla\phi_m - \frac{\partial \mathbf{B}}{\partial t} + \nabla \times \mathbf{A}\right) \quad (19)$$

Now substituting the electric and magnetic field equations into the GDM equations, we obtain two sets of vector and scalar potential fields equations as

$$\frac{1}{c^2}\frac{\partial^2 \mathbf{A}}{\partial t^2} - \nabla^2 \mathbf{A} + \nabla \mathscr{L} - \mu_0 \mathbf{J} = 0 \quad (20)$$

$$\frac{1}{c^2}\frac{\partial^2 \mathbf{B}}{\partial t^2} - \nabla^2 \mathbf{B} + \nabla \mathcal{L} - \mu_0 \mathbf{K} = 0 \quad (21)$$

$$\frac{1}{c^2}\frac{\partial^2 \phi_e}{\partial t^2} - \nabla^2 \phi_e + \frac{\partial L}{\partial t} - \frac{\rho_e}{\varepsilon_0} = 0 \quad (22)$$

$$\frac{1}{c^2}\frac{\partial^2 \phi_m}{\partial t^2} - \nabla^2 \phi_m + \frac{\partial \mathcal{L}}{\partial t} - \mu_0 \rho_m = 0 \quad (23)$$

where $\mathscr{L}$ and $\mathcal{L}$ are represented the relativistic Lorentz gauge conditions, respectively, for electric and magnetic charges of dyons, that is,

$$\mathscr{L} \mapsto \nabla \cdot \mathbf{A} + \frac{1}{c^2}\frac{\partial \phi_e}{\partial t} = 0 \quad (24)$$

$$\mathcal{L} \mapsto \nabla \cdot \mathbf{B} + \frac{1}{c^2}\frac{\partial \phi_m}{\partial t} = 0 \quad (25)$$

Furthermore, in a conducting medium the electromagnetic wave equations for dyons are

$$\frac{1}{c^2}\frac{\partial^2 \mathcal{E}}{\partial t^2} - \nabla^2 \mathcal{E} + \mu_0 \sigma_e \frac{\partial \mathcal{E}}{\partial t} + \nabla\left(\frac{\rho_e}{\varepsilon_0}\right) = 0 \quad (26)$$

$$\frac{1}{c^2}\frac{\partial^2 \mathcal{H}}{\partial t^2} - \nabla^2 \mathcal{H} + \mu_0 \sigma_m \frac{\partial \mathcal{H}}{\partial t} + \nabla(\mu_0 \rho_m) = 0 \quad (27)$$

where $\sigma_e$ and $\sigma_m$ are the conductivities due to electric charge and magnetic monopole, respectively. Thus from (26) and (27), the potential wave propagation for dyons can be expressed as

$$\frac{1}{c^2}\frac{\partial^2 \mathbf{A}}{\partial t^2} - \nabla^2 \mathbf{A} + \mu_0 \sigma_e \frac{\partial \mathbf{A}}{\partial t} + \nabla\left(\nabla \cdot \mathbf{A} + \frac{1}{c^2}\frac{\partial \phi_e}{\partial t} + \mu_0 \sigma \phi_e\right) = 0 \quad (28)$$

$$\frac{1}{c^2}\frac{\partial^2 \mathbf{B}}{\partial t^2} - \nabla^2 \mathbf{B} + \mu_0 \sigma_m \frac{\partial \mathbf{B}}{\partial t} + \nabla\left(\nabla \cdot \mathbf{B} + \frac{1}{c^2}\frac{\partial \phi_m}{\partial t} + \mu_0 \sigma \phi_m\right) = 0 \quad (29)$$

where the Lorentz gauge condition for conducting medium becomes

$$\mathscr{L}^C \mapsto \nabla \cdot \mathbf{A} + \frac{1}{c^2}\frac{\partial \phi_e}{\partial t} + \mu_0 \sigma_e \phi_e = 0 \quad (30)$$

$$\mathcal{L}^C \mapsto \nabla \cdot \mathbf{B} + \frac{1}{c^2}\frac{\partial \phi_m}{\partial t} + \mu_0 \sigma_m \phi_m = 0 \quad (31)$$

If the conductivity of dyons $\sigma(\sigma_e, \sigma_m) \sim 0$, then the Lorentz gauge condition for conducting medium ($\mathscr{L}^C, \mathcal{L}^C$) transfers to the well-known condition ($\mathscr{L}, \mathcal{L}$). All these equations are governed by the classical nature of electromagnetic fields of dyons. In Sects. 3 and 4, we will discuss the quantum analogue of theses classical equations.

## 3. Generalized quantum equations

One of the main open problems of modern physics is that of the foundation of quantum physics. The most basic consequence of quantum physics is that there is a wave associated with the motion of all matter. Thus, starting with the four-dimensional representation of relativistic quantum mechanics, the quaternionic wave function ($\check{\Psi}$) and momentum ($\check{P}$) can be expressed in terms of both vector and scalar components

- four-wave function

$$\check{\Psi} = \left\{\Psi_1, \Psi_2, \Psi_3, \frac{i}{c}\Psi_0\right\} \equiv \left\{\mathbf{\Psi}, \frac{i}{c}\Psi_0\right\} \quad (32)$$

- four-momentum

$$\check{P} = \left\{p_1, p_2, p_3, \frac{i}{c}E\right\} \equiv \left\{\mathbf{p}, \frac{i}{c}E\right\} \quad (33)$$

To discuss the quaternionic quantum mechanics [15], we may identify the quaternionic representation of momentum eigenvalue problem as

$$\breve{P}\breve{\Psi} = m_0 c \breve{\Psi} \tag{34}$$

where $m_0$ denotes the mass of particle (massive dyon). This equation is also governed by the Dirac equation if we use the quaternionic momentum operator $\breve{P} \to i\hbar\gamma^\nu \partial_\nu$, that is, the quaternionic Dirac equation,

$$(i\hbar\gamma^\nu \partial_\nu - m_0 c)\breve{\Psi} = 0 \tag{35}$$

Now from (34), we get

$$\left\{\left(\frac{i}{c}E\right)\boldsymbol{\Psi} + \boldsymbol{p}\left(\frac{i}{c}\Psi_0\right) + \boldsymbol{p} \times \boldsymbol{\Psi}, -\frac{E}{c^2}\Psi_0 - \boldsymbol{p}\cdot\boldsymbol{\Psi}\right\} = m_0 c \left\{\boldsymbol{\Psi}, \frac{i}{c}\Psi_0\right\} \tag{36}$$

Here we have used the quaternionic product. The quaternionic product for two variables $\breve{X}$ and $\breve{Y}$ is

$$\breve{X}\breve{Y} = (x_0 \boldsymbol{y} + \boldsymbol{x} y_0 + \boldsymbol{x} \times \boldsymbol{y}, x_0 y_0 - \boldsymbol{x}\cdot\boldsymbol{y}) \tag{37}$$

Identifying the momentum and energy operators, $\boldsymbol{p} = -i\hbar\nabla$ and $E = i\hbar(\partial/\partial t)$, respectively, we may obtain the following quantum equations:

$$\nabla\cdot\boldsymbol{\Psi} - \frac{1}{c^2}\frac{\partial \Psi_0}{\partial t} - \frac{m_0}{\hbar}\Psi_0 = 0 \tag{38}$$

$$\nabla\Psi_0 - \frac{\partial \boldsymbol{\Psi}}{\partial t} - \frac{m_0 c^2}{\hbar}\boldsymbol{\Psi} = 0 \tag{39}$$

$$\nabla \times \boldsymbol{\Psi} = 0 \tag{40}$$

For second-order quantum equation, operate $\nabla$ on both sides of (38)–(40) and get

$$\nabla^2 \boldsymbol{\Psi} - \frac{1}{c^2}\frac{\partial}{\partial t}\left(\frac{\partial \boldsymbol{\Psi}}{\partial t} + \frac{m_0 c^2}{\hbar}\boldsymbol{\Psi}\right) - \frac{m_0}{\hbar}\left(\frac{\partial \boldsymbol{\Psi}}{\partial t} + \frac{m_0 c^2}{\hbar}\boldsymbol{\Psi}\right) = 0 \tag{41}$$

$$\nabla^2 \Psi_0 - \frac{1}{c^2}\frac{\partial}{\partial t}\left(\frac{\partial \Psi_0}{\partial t} + \frac{m_0 c^2}{\hbar}\Psi_0\right) - \frac{m_0}{\hbar}\left(\frac{\partial \Psi_0}{\partial t} + \frac{m_0 c^2}{\hbar}\Psi_0\right) = 0 \tag{42}$$

along with the identity,

$$\nabla \times (\nabla \times \boldsymbol{\Psi}) \equiv \nabla(\nabla\cdot\boldsymbol{\Psi}) - \nabla^2 \boldsymbol{\Psi} = 0 \tag{43}$$

Equations (41) and (42) reduce to the following compact form:

- vector wave

$$\nabla^2 \boldsymbol{\Psi} - \frac{1}{c^2}\frac{\partial^2 \boldsymbol{\Psi}}{\partial t^2} - 2\left(\frac{m_0}{\hbar}\right)\frac{\partial \boldsymbol{\Psi}}{\partial t} - \frac{m_0^2 c^2}{\hbar^2}\boldsymbol{\Psi} = 0 \tag{44}$$

- scalar wave

$$\nabla^2 \Psi_0 - \frac{1}{c^2}\frac{\partial^2 \Psi_0}{\partial t^2} - 2\left(\frac{m_0}{\hbar}\right)\frac{\partial \Psi_0}{\partial t} - \frac{m_0^2 c^2}{\hbar^2}\Psi_0 = 0 \tag{45}$$

which leads to the quaternionic unified wave

$$\nabla^2 \breve{\Psi} - \frac{1}{c^2}\frac{\partial^2 \breve{\Psi}}{\partial t^2} - 2\left(\frac{m_0}{\hbar}\right)\frac{\partial \breve{\Psi}}{\partial t} - \frac{m_0^2 c^2}{\hbar^2}\breve{\Psi} = 0 \tag{46}$$

Here (44) represents a new form of generalized quantum wave equation for massive vector field ($\boldsymbol{\Psi}$), (45) represents for massive scalar field ($\Psi_0$), whereas (46) describes the quaternionic generalization to the case of unified quantum wave equation. If the mass of the particle is zero ($m_0 \sim 0$), then these quantum wave equations are governed by ordinary wave equations for massless particles, like the photon, graviton, etc., so that $\nabla^2 \breve{\Psi} - (1/c^2)(\partial^2 \breve{\Psi}/\partial t^2) = 0$. Furthermore, (44) and (45) are linear wave equations and hence they satisfy the superposition principle of quantum mechanics. Moreover, we can interpret these equations in terms of generalized Klein–Gorden equations as

$$\left(\Box - \frac{m_0^2 c^2}{\hbar^2}\right)\boldsymbol{\Psi} = \left(\frac{m_0}{\hbar}\right)\frac{\partial \boldsymbol{\Psi}}{\partial t} \tag{47}$$

$$\left(\Box - \frac{m_0^2 c^2}{\hbar^2}\right)\Psi_0 = \left(\frac{m_0}{\hbar}\right)\frac{\partial \Psi_0}{\partial t} \tag{48}$$

where the right-hand part describes the time-dependent source term called the damping terms resulting from the inertia of the massive particle. We can eliminate the damping term by introducing a new time coordinate $\tau$ as

$$\frac{\partial}{\partial \tau} \mapsto \left(\frac{\partial}{\partial t} + \frac{m_0 c^2}{\hbar}\right)$$

then the quantum equations will transform into the following ordinary wave equations:

$$\nabla^2 \boldsymbol{\Psi} - \frac{1}{c^2}\frac{\partial^2 \boldsymbol{\Psi}}{\partial \tau^2} = 0 \quad \Rightarrow \quad \Diamond \boldsymbol{\Psi} = 0 \tag{49}$$

$$\nabla^2 \Psi_0 - \frac{1}{c^2}\frac{\partial^2 \Psi_0}{\partial \tau^2} = 0 \quad \Rightarrow \quad \Diamond \Psi_0 = 0 \tag{50}$$

where $\Diamond = [\nabla^2 - (1/c^2)(\partial^2/\partial \tau^2)]$ denotes new d'Alembert operator for time coordinate $\tau$. Now, if we consider a general plane wave solution of (46) as

$$\breve{\Psi} = \xi \exp i(\omega t - \boldsymbol{k}\cdot\boldsymbol{r}) \quad \text{where } \xi = \text{constant} \tag{51}$$

then the dispersion relation can be expressed as

$$\omega^2 - \left(2i\frac{m_0 c^2}{\hbar}\right)\omega - \left(\frac{m_0^2 c^4}{\hbar^2} + c^2 k^2\right) = 0 \tag{52}$$

There are two solutions of (52): Dirac positive and negative solutions (i.e., positive energy with particle energy $\hbar\omega^+$ and negative energy with antiparticle energy $\hbar\omega^-$). Thus,

$$\hbar^2 |\omega^\pm|^2 = m_0^2 c^4 + c^2 k^2 \hbar^2 \quad \Rightarrow \quad E^2 = m_0^2 c^4 + p^2 c^2 \tag{53}$$

Equation (53) shows the eigenvalue relation. The interpretation of (53) resembles the Dirac theory of antiparticle. Moreover the group and phase velocities, $v_g$ and $v_p$, respectively, can be written as

$$v_g = \frac{\partial \omega^\pm}{\partial k} = \pm c \tag{54}$$

$$v_p = \frac{\omega^\pm}{k} = \frac{im_0c^2}{\hbar k} \pm c \tag{55}$$

If we consider a particle with $m_0 = 0$ and propagate a quantum wave, then its group and phase velocity will be the same and equal to the speed of light.

## 4. Quantum structure for the electromagnetic wave propagation of dyons

In the previous section we have obtained the quantum field equations for massive particles. In the electromagnetic case, Dirac [7, 9] studied the problem of the quantum mechanics of a particle in the presence of a magnetic monopole and found that a consistent quantization forced a relation between the electric charge of the particle and the magnetic charge of the monopole called the Dirac quantization condition. For the interaction of two particles having electric and magnetic charges $(e_1, g_1)$ and $(e_2, g_2)$, Schwinger [16] extended the Dirac quantization condition $eg = (1/2)n$, quantum number $n = 1, 2, 3, \ldots$, for the case of dyons, that is,

$$(e_1g_2 - e_2g_1) = \frac{1}{2}n \tag{56}$$

Furthermore, the Witten effect [17], demonstrated by Edward Witten, states that the electric charges of dyons must all be equal, modulo one, to the product of their magnetic charge and the theta-angle of the theory. On the other hand, if we do not consider Dirac particles as dyons, the Dirac quantization condition loses its dual invariance. Thus dyon plays an important role in electromagnetic duality with the association of Chirality quantization parameter and it is important to consider the consistent quantum field theory for the simultaneous existence of electric and magnetic charges. Now it is also expected that it appears just as the energy density of generalized electromagnetic-vector field, thus we may write the mass function $f(m_0)$ for dyons as

$$\text{Mass (dyon)} = f(m_0) \sim f\left[\sqrt{(e^2 + g^2)}\right] \tag{57}$$

which plays an important role in supersymmetric gauge theories. Correspondingly, Bogomolny bound [18] shows the mass of an individual dyon, given by

$$m_0 = u|e + ig| = u\sqrt{(e^2 + g^2)} \tag{58}$$

where $e$ is electric charge, $g$ is charge of monopole, and $u$ is the magnitude of the vacuum expectation value of scalar Higgs field. In Higgs mechanism, the Higgs field provides the mechanism whereby the vacuum spontaneously breaks the SU(2) gauge symmetry down to the U(1) group.

Thus to develop the relativistic quantum theory [19, 20] of electromagnetic wave propagation of dyons, we compare (20) and (44) and get the following quantum equation:

$$\frac{1}{c^2}\frac{\partial^2 \mathbf{A}}{\partial t^2} - \nabla^2 \mathbf{A} + 2\left(\frac{m_0}{\hbar}\right)\frac{\partial \mathbf{A}}{\partial t} + \frac{m_0^2 c^2}{\hbar^2}\mathbf{A} = 0 \tag{59}$$

where (59) represents a relativistic quantum wave equation of an electron for electric vector potential ($\mathbf{A}$); here we have used the following identities:

$$\mathcal{L} \mapsto \left(\nabla \cdot \mathbf{A} + \frac{1}{c^2}\frac{\partial \phi_e}{\partial t}\right) = -\frac{2m_0\phi_e}{\hbar} \tag{60}$$

$$\nabla \phi_e = -\frac{\partial \mathbf{A}}{\partial t} \tag{61}$$

$$\mathbf{J} = -\left(\frac{m_0^2 c^2}{\mu_0 \hbar^2}\right)\mathbf{A} \tag{62}$$

Equation (60) is denoted the generalized Lorentz gauge condition for electric potential of massive dyons and (61) is denoted the strength of the longitudinal component of electric field vector ($\mathcal{E}_l \sim \nabla \phi_e$), while (62) is denoted the quantized electric current density connected with electric vector potential. From (61) and (62), we also can express that the rate of change of electric current density depends on the longitudinal component of electric field (i.e., $(\partial \mathbf{J}/\partial t) = (m_0^2 c^2/\mu_0 \hbar^2)\mathcal{E}_l$). Therefore, if $\mathcal{E}_l \to 0$, we obtain the conserved electric current density, called Noether current. As such, comparing (21) and (44), we investigate the quantum wave equation for magnetic vector potential ($\mathbf{B}$), that is,

$$\frac{1}{c^2}\frac{\partial^2 \mathbf{B}}{\partial t^2} - \nabla^2 \mathbf{B} + 2\left(\frac{m_0}{\hbar}\right)\frac{\partial \mathbf{B}}{\partial t} + \frac{m_0^2 c^2}{\hbar^2}\mathbf{B} = 0 \tag{63}$$

with

$$\mathcal{L} \mapsto \left(\nabla \cdot \mathbf{B} + \frac{1}{c^2}\frac{\partial \phi_m}{\partial t}\right) = -\frac{2m_0\phi_m}{\hbar} \tag{64}$$

$$\nabla \phi_m = -\frac{\partial \mathbf{B}}{\partial t} \tag{65}$$

$$\mathbf{K} = -\left(\frac{m_0^2 c^2}{\mu_0 \hbar^2}\right)\mathbf{B} \tag{66}$$

Here, (63) represents the quantum wave equation for magnetic monopole. The generalized Lorentz gauge condition for massive monopole is given by (64). As such, (65) represents the longitudinal component of magnetic field vector ($\mathcal{H}_l \sim \nabla \phi_m$) and (66) denotes the quantized magnetic current density connected with magnetic vector potential of dyons. Furthermore, we can obtain the magnetic Noether's conserved current from (65) and (66) if $\mathcal{H}_l \to 0$. Generalized gauge equations (60) and (64) are invariant under gauge transformation, that is, $\mathbf{A}' \to \mathbf{A} + \nabla \Lambda$, $\phi'_e \to \phi_e - (\partial \Lambda/\partial t)$ and $\mathbf{B}' \to \mathbf{B} + \nabla \Lambda$, $\phi'_m \to \phi_m - (\partial \Lambda/\partial t)$, where $\Lambda$ satisfy the following relation:

$$\frac{1}{c^2}\frac{\partial^2 \Lambda}{\partial t^2} - \nabla^2 \Lambda + 2\left(\frac{m_0}{\hbar}\right)\frac{\partial \Lambda}{\partial t} = 0 \tag{67}$$

The solution of (67) can be written as $\Lambda = \Upsilon \exp(-mc^2t/\hbar)$, where $\Upsilon$ satisfies the Klein–Gordon equation. Similarly, comparing scalar equations (22), (45) and (23), (45), we obtain the following quantum wave equations for dyonic scalar potentials ($\phi_e, \phi_m$):

$$\frac{1}{c^2}\frac{\partial^2 \phi_e}{\partial t^2} - \nabla^2 \phi_e + 2\left(\frac{m_0}{\hbar}\right)\frac{\partial \phi_e}{\partial t} + \frac{m_0^2 c^2}{\hbar^2}\phi_e = 0 \tag{68}$$

and

$$\frac{1}{c^2}\frac{\partial^2 \phi_m}{\partial t^2} - \nabla^2 \phi_m + 2\left(\frac{m_0}{\hbar}\right)\frac{\partial \phi_m}{\partial t} + \frac{m_0^2 c^2}{\hbar^2}\phi_m = 0 \tag{69}$$

along with the following relation of charge densities:

- electric charge density

$$\rho_e = -\left(\frac{m_0^2}{\mu_0 \hbar^2}\right)\phi_e \tag{70}$$

- magnetic charge density

$$\rho_m = -\left(\frac{m_0^2 c^2}{\mu_0 \hbar^2}\right)\phi_m \tag{71}$$

We should note that the quantum analog of electric charge density is connected to the electric scalar potential, whereas the magnetic charge density is connected with the magnetic scalar potential of dyons. If we introduce $\lambda = h/m_0 c$ as the Compton wavelength for dyons, then the propagation vector ($k$) of dyonic wave can be defined by $k^2 = (4\pi^2/\lambda^2) \simeq (m_0^2 c^2/\hbar^2)$. Finally, we can relate the quantum structure of four-current density in terms of generalized four-potential of dyons (i.e., $(J, K) = -(1/\mu_0)k^2(A, B)$; $(\rho_e, \rho_m) = -(1/\mu_0)k^2(\phi_e/c^2, \phi_m)$). Correspondingly, the wave propagation of four-current $\{S, \varphi\}$ can be transformed as the following differential equations:

$$\frac{1}{c^2}\frac{\partial^2 S}{\partial t^2} - \nabla^2 S + 2\left(\frac{m_0}{\hbar}\right)\frac{\partial S}{\partial t} + \frac{m_0^2 c^2}{\hbar^2}S = 0 \quad (\forall\ S \mapsto (J, K)) \tag{72}$$

$$\frac{1}{c^2}\frac{\partial^2 \varphi}{\partial t^2} - \nabla^2 \varphi + 2\left(\frac{m_0}{\hbar}\right)\frac{\partial \varphi}{\partial t} + \frac{m_0^2 c^2}{\hbar^2}\varphi = 0 \quad (\forall\ \varphi \mapsto (\phi_e, \phi_m)) \tag{73}$$

where (72) and (73) are describe the unified quantum wave propagation in terms of current and charge densities of dyons. Now, substituting the value of ($A$, $\phi_e$) and ($B$, $\phi_m$) in (60) and (64), we obtain the new interesting results for massive dyons, that is, say,

$$\nabla \cdot J + \frac{\partial \rho_e}{\partial t} = -\frac{2m_0 c^2}{\hbar}\rho_e = \mathcal{T}_e \tag{74}$$

$$\nabla \cdot K + \frac{1}{c^2}\frac{\partial \rho_m}{\partial t} = -\frac{2m_0}{\hbar}\rho_m = \mathcal{T}_m \tag{75}$$

These equations represent the generalized continuity equations for massive dyons in the case of quantum fields. Here ($\mathcal{T}_e$, $\mathcal{T}_m$) shows the torque density resulting from the two spin states. If producing a polarization [21], $P^s$ is given by $\mathcal{T}_{e,m} = -\nabla \cdot P^s_{e,m}$. Hence, the generalized continuity equations transform as

$$\nabla \cdot (J + P^s_e) + \frac{\partial \rho_e}{\partial t} = 0 \mapsto \nabla \cdot J_{\text{effective}} + \frac{\partial \rho_e}{\partial t} = 0 \tag{76}$$

$$\nabla \cdot (K + P^s_m) + \frac{1}{c^2}\frac{\partial \rho_m}{\partial t} = 0 \mapsto \nabla \cdot K_{\text{effective}} + \frac{1}{c^2}\frac{\partial \rho_m}{\partial t} = 0 \tag{77}$$

where the effective current density for electric and magnetic charge of dyons are defined as, $J_{\text{effective}} = J + P^s_e$ and $K_{\text{effective}} = K + P^s_m$, respectively. Moreover, for conducting medium, the quantum analogue of electric and magnetic wave propagation of dyons can be expressed as

$$\frac{1}{c^2}\frac{\partial^2 \mathcal{E}}{\partial t^2} - \nabla^2 \mathcal{E} + 2\left(\frac{m_0}{\hbar}\right)\frac{\partial \mathcal{E}}{\partial t} + \frac{m_0^2 c^2}{\hbar^2}\mathcal{E} = 0 \tag{78}$$

$$\frac{1}{c^2}\frac{\partial^2 \mathcal{H}}{\partial t^2} - \nabla^2 \mathcal{H} + 2\left(\frac{m_0}{\hbar}\right)\frac{\partial \mathcal{H}}{\partial t} + \frac{m_0^2 c^2}{\hbar^2}\mathcal{H} = 0 \tag{79}$$

Here we assume that the quantized dual conductivity become almost the same (i.e., $\sigma_e \sim \sigma_m = 2m_0/\hbar\mu_0$) and the dual quantized fields become $\mathcal{E} = (\hbar^2/m_0^2 c^2)(1/\varepsilon_0)(\nabla \rho_e)$ and $\mathcal{H} = (\hbar^2/m_0^2 c^2)\mu_0(\nabla \rho_m)$. Accordingly, substituting the value of $\rho_e$ and $\rho_m$ from (70) and (71), we also get the well-known relation of longitudinal components of electric and magnetic field. Furthermore, we conclude that the wave propagation of dyons in a medium has the compact quantum wave structure given by $(\Box - k^2)\Omega = (2k/c)(\partial \Omega/\partial t)$, where the time-varying term $\partial \Omega/\partial t$ can be seen as a dissipative (damping) term resulting from the inertia of the dyons, or the motion of the dyons in space–time.

## 5. Quantum wave propagation in an external electromagnetic field of dyons

To describe the quantum wave equations for a four-dimensional particle dyon in an external electromagnetic field, the quantum operators of energy and momentum can be written as

$$\check{P} = (\check{P} + Q\check{V}) = \begin{cases} p \mapsto p + QV \\ E \mapsto E + Q\Phi \end{cases} \tag{80}$$

where $V \to (A + iB)$ and $\Phi \to (\phi_e + i\phi_m)$ are vector and scalar potentials of the dyon [22]. Moreover, the classical field equations of dyons and its conservation laws have already been discussed [23–27]. From (80), we get the following changes in differential operators:

$$\nabla \mapsto \left(\nabla - \frac{Q}{i\hbar}V\right) \qquad \frac{\partial}{\partial t} \mapsto \left(\frac{\partial}{\partial t} + \frac{Q}{i\hbar}\Phi\right) \tag{81}$$

Now, applying the preceding space–time operators in quaternionic momentum eigenvalue problem (34) and obtain the following relativistic field equations:

- imaginary part of the scalar field

$$\nabla \cdot \Psi - \frac{1}{c^2}\frac{\partial \Psi_0}{\partial t} - \frac{m_0}{\hbar}\Psi_0 = 0 \tag{82}$$

- real part of the vector field

$$\nabla \Psi_0 - \frac{\partial \Psi}{\partial t} + \frac{Qc}{\hbar}(V \times \Psi) - \frac{m_0 c^2}{\hbar}\Psi = 0 \tag{83}$$

- imaginary part of the vector field

$$\nabla \times \Psi - \frac{Qc}{\hbar}(\Phi \Psi + V\Psi_0) = 0 \tag{84}$$

- imaginary part of the scalar field

$$\Phi \Psi_0 + c^2(V \cdot \Psi) = 0 \tag{85}$$

**Table 1.** Quantum equations for electromagnetic fields of dyons.

| London equations | $\frac{\partial \boldsymbol{J}_s}{\partial t} = \left(\frac{n_s e^2}{m_e}\right)\boldsymbol{\mathcal{E}}$ | $\nabla \times \boldsymbol{J}_s = -\left(\frac{n_s e^2}{m_e}\right)\boldsymbol{\mathcal{H}}$ | $\boldsymbol{J}_s = -\left(\frac{n_s e^2}{m_e}\right)\boldsymbol{A}$ |
|---|---|---|---|
| Quantum analog of London's equations for dyons | $\frac{\partial \boldsymbol{J}}{\partial t} = \left(\frac{m_0^2 c^2}{\mu_0 \hbar^2}\right)\boldsymbol{\mathcal{E}}_l$ | $\nabla \times \boldsymbol{J} = -\left(\frac{m_0^2 c^2}{\mu_0 \hbar^2}\right)\boldsymbol{\mathcal{H}}$ | $\boldsymbol{J} = -\left(\frac{m_0^2 c^2}{\mu_0 \hbar^2}\right)\boldsymbol{A}$ |
| | $\frac{\partial \boldsymbol{K}}{\partial t} = \left(\frac{m_0^2 c^2}{\mu_0 \hbar^2}\right)\boldsymbol{\mathcal{H}}_l$ | $\nabla \times \boldsymbol{K} = -\left(\frac{m_0^2 c^2}{\mu_0 \hbar^2}\right)\boldsymbol{\mathcal{E}}$ | $\boldsymbol{K} = -\left(\frac{m_0^2 c^2}{\mu_0 \hbar^2}\right)\boldsymbol{B}$ |

**Note:** $J_s$ is the superconducting current density, $e$ is the charge of an electron and proton, $m_e$ is electron mass, and $n_s$ is a phenomenological constant loosely associated with a number density of superconducting carriers.

Here we can interpret these quantum equations by a wave equations of dyons in external fields if we operate $\nabla$ to (82) and (83) and also use (84) and (85). Thus the generalized quantum wave equations become

$$\nabla^2 \boldsymbol{\Psi} - \frac{1}{c^2}\frac{\partial^2 \boldsymbol{\Psi}}{\partial t^2} + \frac{Q}{\hbar c}\frac{\partial \boldsymbol{\Gamma}}{\partial t} - 2\left(\frac{m_0}{\hbar}\right)\frac{\partial \boldsymbol{\Psi}}{\partial t} + m_0\frac{Qc}{\hbar^2}\boldsymbol{\Gamma} - \frac{m_0^2 c^2}{\hbar^2}\boldsymbol{\Psi} = 0 \quad (86)$$

$$\nabla^2 \Psi_0 - \frac{1}{c^2}\frac{\partial^2 \Psi_0}{\partial t^2} + \frac{Qc}{\hbar}\nabla \cdot \boldsymbol{\Gamma} - 2\left(\frac{m_0}{\hbar}\right)\frac{\partial \Psi_0}{\partial t} - \frac{m_0^2 c^2}{\hbar^2}\Psi_0 = 0 \quad (87)$$

These equations show the generalized wave equations of quaternionic quantum state ($\boldsymbol{\Psi}$, $\Psi_0$), where (86) represents the vector-field wave equation and (87) represents the scalar-field wave equation for massive dyons in external fields. The term $\boldsymbol{\Gamma} = (V \times \boldsymbol{\Psi})$ describes the potential interaction to quantum state $\boldsymbol{\Psi}$, which may provide electromagnetic field of dyons. The compact form of quantum wave equations with external fields $\boldsymbol{\Gamma}$ can be written as

$$\nabla^2 \boldsymbol{\Psi} - \frac{1}{c^2}\frac{\partial^2 \boldsymbol{\Psi}}{\partial \tau^2} + \frac{Q}{\hbar c}\frac{\partial \boldsymbol{\Gamma}}{\partial \tau} = 0 \quad (88)$$

$$\nabla^2 \Psi_0 - \frac{1}{c^2}\frac{\partial^2 \Psi_0}{\partial \tau^2} + \frac{Qc}{\hbar}(\nabla \cdot \boldsymbol{\Gamma}) = 0 \quad (89)$$

In the absence of field interaction ($\boldsymbol{\Gamma}$), (88) and (89) are represented to the ordinary wave equations for quaternionic quantum state. In (89), the divergence of $\boldsymbol{\Gamma}$ can be expressed as

$$\nabla \cdot \boldsymbol{\Gamma} = \boldsymbol{\Psi} \cdot (\nabla \times V) - V \cdot (\nabla \times \boldsymbol{\Psi}) \quad (90)$$

whereas

$$V \cdot (\nabla \times \boldsymbol{\Psi}) = \frac{Qc}{\hbar}\left(V^2 - \frac{\Phi^2}{c^2}\right)\Psi_0 \quad (91)$$

and

$$\boldsymbol{\Psi} \cdot (\nabla \times V) = i(\boldsymbol{\Psi} \cdot \boldsymbol{\Omega}) + \boldsymbol{\Psi} \cdot (\nabla \Phi) + \boldsymbol{\Psi} \cdot \left(\frac{\partial V}{\partial t}\right) \quad (92)$$

In (92), the first term on the right-hand side is associated with the generalized electromagnetic field $\boldsymbol{\Omega}(\boldsymbol{\mathcal{E}}, \boldsymbol{\mathcal{H}})$ of dyons. The scalar wave propagation (87) and (89) behave as a matter wave, which propagates with scalar electromagnetic field term $\boldsymbol{\Psi} \cdot \boldsymbol{\Omega}$. Thus the scalar field equation represents the equation of motion of a quantum scalar or pseudo-scalar field, a field whose quanta are spinless particles. Similarly, the solution of vector wave propagation (86) provides the Dirac-like solution of fermiomic particles. Any solution to the Dirac equation is automatically a solution to the Klein–Gordon equation, but the converse is not true.

## 6. Conclusion

In the present paper, we have discussed the quaternionic quantum wave equations for dyons. The advantage of quaternion formulation is that it shows a four-dimensional representation and can be transformed in a specific way under Lorentz transformations. We have obtained the first- and second-order differential quantum equation of quaternionic state $\tilde{\Psi}$. The quaternionic unification of quantum wave equations have been expressed in a simple and consistent manner. Keeping in mind the quaternionic quantum formulation, we have investigated the relativistic quantum theory of electromagnetic wave propagation of dyons. This analogous quantum theory leads to a generalized Lorentz gauge condition for the electric charge and magnetic monopoles. The beauty of this quantum formulation is that the quantum analog of electric charge density is connected to the electric scalar potential, whereas the magnetic charge density is connected with the magnetic scalar potential of dyons. Equations (74) and (75) are the generalized continuity equations for massive dyons in the presence of quantum fields. We have described the quantum theory of electric and magnetic wave propagation of dyons in terms of the quaternionic formulation. Moreover, in the theory of superconductivity the analogous representation of London equations [28] in terms of dyons can be related by our unified quantum theory of electrodynamics. Using (61), (62) and (65), (66), we summarized the generalized London equations for dyons in Table 1.

Thus the major triumph of these equations are their ability to explain the Meissner effect [29] wherein a material exponentially expels all internal magnetic fields as it crosses the superconducting threshold. Therefore, the quantum formulation of electromagnetic field of dyons can be used to explain the properties of superconducting materials.